\documentclass[12pt]{article}
\usepackage{amsmath,amsfonts,amssymb,amsthm}

\usepackage{amsthm}
\usepackage[a4paper=true]{hyperref}

\voffset = - 1.2cm
\begin{document}

\centerline{\Large \bf Miura-Reciprocal transformations for two integrable} 
 \centerline{\Large\bf hierarchies in $1+1$ dimensions} \vskip 0.25cm
\centerline{P.G. Est\'evez and C.
Sard\'on}
\medskip
\vskip 0.5cm
\centerline{Department of Fundamental Physics, University of
Salamanca.}
\medskip
\centerline{Plza. de la Merced s/n, 37.008, Salamanca, Spain.}
\medskip
\centerline{\it{cristinasardon@usal.es}}
\vskip 1cm

\abstract{We here present two different hierarchies of PDEs in $1+1$ dimensions whose first and second member are the shallow water wave Camassa-Holm and Qiao
equations, correspondingly. These two hierarchies can be transformed by reciprocal methods into the Calogero-Bogoyanlevski-Schiff equation (CBS) and its modified version (mCBS), respectively.
Considering that there exists a Miura transformation between the CBS and mCBS, we shall obtain a relation between the initial hierarchies
by means of a composition of a Miura and reciprocal transforms.}

\section{Introduction}

Reciprocal methods are based on transformations in which {\it the role of the independent and dependent variables
is interchanged}. When the variables are switched, the space of independent variables is called the {\it reciprocal space}, or in the case of two dimensions, the
{\it reciprocal plane}. As a physical interpretation, whereas the independent variables play the usual role of positions,
in the reciprocal space, this number is increased by turning certain fields (usually velocities or parametrized heights of waves, in the case
of fluid mechanics) into independent variables and vice-versa \cite{Sardon:Conte08}.

\medskip

Lately, a lot of attention has been laid upon reciprocal transformations, as they conform a very useful instrument for the identification of ODEs, PDEs or high-order hierarchies which, a priori, do not
have the Painlev\'e property (PP) \cite{Sardon:estevez09}, \cite{Sardon:estevez04}, \cite{Sardon:estevez05-1}.

According to the Painlev\'e criteria for integrability, we say that a non-linear equation is integrable if its solutions are single-valued
in the neighborhood of the movable singularity manifold.
The PP can be checked through means of an algorithmic procedure developed by Weiss \cite{Sardon:weiss83} ,\cite{Sardon:wtc}, which gives us a set of solutions relying on the truncation of an
infinite Laurent expansion. In this situation, one can prove that the manifold of movable singularities satisfies a set of equations
known as {\it the singular manifold equations} (SME).
Also, the SME are a subject of our rising interest due to the possibility of classification of O/PDEs in terms of their SME used as a canonical
form. If this were the case, two apparently unrelated equations sharing the same SME must be tantamount versions of a unique equation.
In this way, we are reducing the ostensible bundle of non-linear integrable equations into series of equivalent ones.

\medskip

The PP is non-invariant under changes of independent/dependent variables. For this reason, we can transform an initial equation
which does not pass the Painlev\'e test into another which successfully accomplishes the Painlev\'e requirements. In most of the cases,
the transformed equation is well-known, as well as its properties, integratibility and so on. In this way, reciprocal transformations can be very useful, for instance, in the derivation of Lax pairs (LP) in such a way
that given the LP for the transformed equation, we shall perform the inverse reciprocal transformation to obtain the LP for the initial.

\medskip

Nevertheless, finding a suitable reciprocal transform is usually a complicated task and of very limited use. Notwithstanding, in cases
of fluid mechanics, a change of this type is usually reliable. In \cite{Sardon:DHH02} ,\cite{Sardon:h00}, similar transformations were introduced to turn peakon
equations into other integrable ones.

Sometimes, the composition of a Miura and a reciprocal transform, hence the name {\it Miura-reciprocal transform},
helps us to relate two different hierarchies, which is the purpose of the present paper. We shall illustrate this matter
through a particular example: the Camassa-Holm and Qiao hierarchies in $1+1$ dimensions. We shall prove that a combination of a Miura and reciprocal transforms
can relate the parametrized height of the wave in Camassa-Holm's $U=U(X,T)$ (CH$1+1$) with Qiao's $u=u(x,t)$ (Qiao$1+1$) hierarchies.
A summary of the process is included within the following diagram:

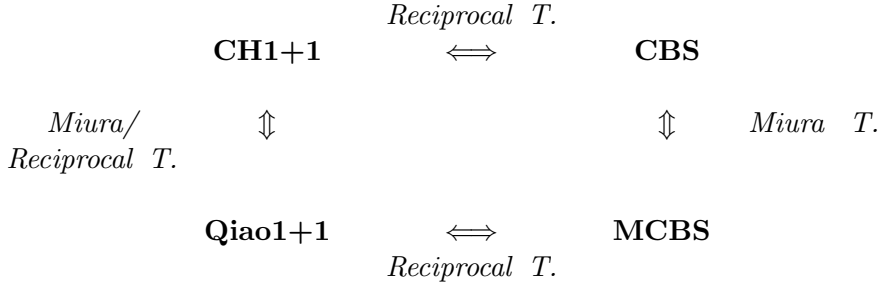
\begin{eqnarray}\nonumber
\begin{array} {ccccc}&
& \quad\textit{\textrm{Reciprocal\, T.}}\quad & &\\
&\textbf{\textrm{CH}1+1} & \Longleftrightarrow &\textbf{ \textrm{CBS }}& \\ & & &
&\\\textit{\textrm{Miura/}}& \Updownarrow & & \Updownarrow &
\textit{\textrm{Miura \, T.}}\\ \textit{\textrm{Reciprocal\,  T.}}& & & &\\ & & & & \\
 &\textbf{\textrm{Qiao}1+1} & \Longleftrightarrow & \textbf{\textrm{MCBS }}&\\ &
&\quad \textit{\textrm{Reciprocal \,T.}}\quad & &
\end{array}
\end{eqnarray}

In view of this, we shall present the plan of the paper as follows:
In Section 2, we establish the reciprocal transformations that link CH$1+1$ and Qiao$1+1$ to CBS and mCBS, respectively.
Section 3 is devoted to the introduction of the Miura transformation between CBS and mCBS, as well as the inverse reciprocal transform
from CBS and mCBS to the initial CH$1+1$ and Qiao$1+1.$
In closing, we underline several important relations among the fields and independent variables present in both initial hierarchies.

\section{Reciprocal Links}

In this section we aim to review previous results from one of us \cite{Sardon:estevez51} ,\cite{Sardon:estevez05-1}, in which the reciprocal links for the Camassa-Holm
and Qiao hierarchies in $2+1$ dimensions were introduced, and particularize these to the case of $1+1$ dimensions that concerns us in this letter.
Notice that technical details will be omitted but can be checked in the aforemention references. It is important to mention that
henceforth we shall use capital letters for the independent variables and fields present in the Camassa Holm's hierarchy and lower case letters
for those variables and fields concerning Qiao's.

\subsection{Reciprocal link for CH$1+1$}

The n-component Camassa-Holm hierarchy in $1+1$ dimensions can be written in a compact form in terms of a recursion operator as follows:
\begin{equation}\nonumber
U_T=R^{-n}U_X.
\end{equation}
with $K,J$ defined as $K=\partial_{XXX}-\partial_{X}$ and $J=-\frac{1}{2}(\partial_XU+U\partial_X)$. The factor $-\frac{1}{2}$ has
been conveniently added for future calculations.
If we use that $R=KJ^{-1}$ and include auxiliary fields $\Omega^{(i)}$ with $i=1,\dots,n$ when
the inverse of an operator appears,
\begin{align}
U_T&=J\Omega^{(1)},\nonumber\\
K\Omega^{(i)}&=J\Omega^{(i+1)},\quad i=1,\dots,n-1,\label{chh}\\
U_X&=K\Omega^{(n)}\nonumber
\end{align}

It is also useful to introduce the change $U=P^2$ such that the final equations read:
\begin{align}
P_T&=-\frac{1}{2}\left(P\Omega^{(1)}\right)_X, \label{chcf}\\
\Omega^{(i)}_{XXX}-\Omega^{(i)}_{X}&=-P\left(P\Omega^{(i+1)}\right)_X,\quad i=1,\dots,n-1\label{cht1}\\
P^2&=\Omega^{(n)}_{XX}-\Omega^{(n)}.\label{cht2}
\end{align}
Given the conservative form of equation \eqref{chcf}, the following transformation arises naturally:
\begin{equation}\label{rtch}
dT_0=PdX-\frac{1}{2}P\Omega^{(1)}dT, \quad dT_1=dT
\end{equation}
such that $d^{2}T_0=0$, recovering \eqref{chcf}.
We shall now propose a reciprocal transformation \cite{Sardon:DHH02} by considering the former independent variable $X$ as a dependent field
of the new pair of independent variables $X=X(T_0,T_1),$ and therefore, $dX=X_0\,dT_0+X_1\,dT_1$ where the subscripts
zero and one refer to partial derivative of the field $X$ with respect to $T_0$ and $T_1$, correspondingly.
The inverse transformation takes the form:
\begin{equation}\label{irtch}
dX=\frac{dT_0}{P}+\frac{1}{2}\Omega^{(1)}dT_1,\quad dT=dT_1
\end{equation}
which, by direct comparison with the total derivative of the field $X$, we obtain:
\begin{equation}
X_0=\frac{1}{P},\quad X_1=\frac{\Omega^{(1)}}{2}.
\end{equation}
We can now extend the transformation \cite{Sardon:estevez05-1} by introducing $n-1$ independent variables $T_2,\dots,T_n$ which account for the transformation
of the auxiliary fields $\Omega^{(i)}$ in such a way that $\Omega^{(i)}=2X_i$, with $i=2,\dots,n$ and $X_i=\frac{\partial X}{\partial T_i}.$ Then, $X$ is a function
 $X=X(T_0,T_1,T_2,\dots,T_n)$ of $n+1$ variables.
It requires some computation to transform the hierarchy (2)-(4) into the equations that  $X=X(T_0,T_1,T_2,\dots,T_n)$ should obey. For this matter, we use the symbolic calculus package Maple.
Equation \eqref{chcf} is identically satisfied by the transformation and \eqref{cht1}, \eqref{cht2} lead to the following
set of PDEs:
\begin{equation}\label{bcbs}
-\left(\frac{X_{i+1}}{X_0}\right)_0=\left\{\left(\frac{X_{00}}{X_0}+X_0\right)_0-\frac{1}{2}\left(\frac{X_{00}}{X_0}+X_0\right)^2\right\}_i,\quad i=1,\dots,n-1.
\end{equation}
which constitutes $n-1$ copies of the same system, each of which is written in three variables $T_0,T_i,T_{i+1}$.
Considering the conservative form of \eqref{bcbs}, we shall introduce the change:
\begin{align}
M_i&=-\frac{1}{4}\left(\frac{X_{i+1}}{X_0}\right),\\
M_0&=\frac{1}{4}\left\{\left(\frac{X_{00}}{X_0}+X_0\right)_0-\frac{1}{2}\left(\frac{X_{00}}{X_0}+X_0\right)^2\right\}\label{cbs2}
\end{align}
with $M=M(T_0,T_i,T_{i+1})$ and $i=1,\dots,n-1$.
The compatibility condition of $X_{000}$ and $X_{i+1}$ in this system gives rise to a set of  equations written entirely in terms of $M$:
\begin{equation}\label{cbs}
M_{0,i+1}+M_{000i}+4M_iM_{00}+8M_0M_{0i}=0,\quad i=1,\dots,n-1
\end{equation}
that are  $n-1$ CBS equations \cite{Sardon:cbs}, \cite{Sardon:cal} each one in three variables $M=M(T_0,..T_i,T_{i+1},...T_n)$. These equations have the PP \cite{Sardon:pick}, \cite{Sardon:painleve} and the singular manifold method (SMM) can be
applied to derive its LP. Making use of CBS's LP and with the aid of the inverse reciprocal transform,
we could derive a LP for the initial CH$1+1$. The detailed process can be consulted in \cite{Sardon:estevez04}, \cite{Sardon:estevez05-1}.

\subsection{Reciprocal link for Qiao$1+1$}

The $n$-component Qiao hierarchy in $1+1$ dimensions can be written in a compact form in terms of a recursion operator:
\begin{equation}
u_t=r^{-n}u_x \nonumber
\end{equation}
such that $r=kj^{-1}$ and  $k=\partial_{xxx}-\partial_{x}$ and $j=-\partial_xu(\partial_x)^{-1}u\partial_x$.. If we introduce $n$ additional fields $v^{(i)}$ when
we encounter the inverse of an operator, the expanded equations read:
\begin{align}
u_t&=jv^{(1)},\nonumber\\
kv^{(j)}&=jv^{(i+1)},\quad i=1,\dots,n-1,\label{qh}\\
u_x&=kv^{(n)},\nonumber
\end{align}

This hierarchy was first introduced by one of us in \cite{Sardon:estevez51} as a generalization of the Qiao hierarchy to $2+1$ dim., depicted
in \cite{Sardon:qiao2007} and whose second member was studied in \cite{Sardon:qiao}.
If we now introduce the value of the operators $k$ and $j$, we obtain the following equations:
\begin{align}
u_t&=-\left(u\omega^{(1)}\right)_x,\label{qcf}\\
v^{(i)}_{xxx}-v^{(i)}_x&=-\left(u\omega^{(i+1)}\right)_x,\quad i=1,\dots,n-1,\label{qt1}\\
u&=v^{(n)}_{xx}-v^{(n)}.\label{qt2}
\end{align}
in which the change $\omega^{(i)}_x=uv^{(i)}_x$ with other $n$ auxiliary fields $\omega^{(i)}$ has necessarily been included to operate with the inverse term present in $j$.

Given the conservative form of \eqref{qcf}, the following change reciprocal transformation \cite{Sardon:DHH02} naturally:
\begin{equation}\label{rtq}
dT_0=u\,dx-u\omega^{(1)}dt,\quad dT_1=dt
\end{equation}
such that $d^{2}T_0=0$ recovers \eqref{qcf}. We now propose a reciprocal transformation \cite{Sardon:estevez05-1} by considering the initial independent
variable $x$ as a dependent field of the new independent variables such that $x=x(T_0,T_1)$, and therefore, $dx=x_0\,dT_0+x_1\,dT_1$.
The inverse transformation adopts the form:
\begin{equation}\label{irtq}
dx=\frac{dT_0}{u}+\omega^{(1)}dT_1,\quad dt=dT_1.
\end{equation}
By direct comparison of the inverse transform with the total derivative of $x$, we obtain that:
\begin{equation}
x_0=\frac{1}{u},\quad x_1=\omega^{(1)}.
\end{equation}
We shall prolong this transformation in such a way that we introduce new variables $T_2,\dots,T_n$ such that $x=x(T_0, T_1,...T_n)$ according to the following rule $\omega^{(i)}=x_i,  x_i=\frac{\partial x}{\partial T_i}$ for $i=2,\dots,n$.
In this way, \eqref{qcf} is identically satisfied by the transformation and \eqref{qt1}, \eqref{qt2} are transformed into $n-1$ copies
of the following equation, which is written in terms of three variables $T_0,T_i,T_{i+1}$:
\begin{equation}\label{bmcbs}
\left(\frac{x_{i+1}}{x_0}+\frac{x_{i00}}{x_0}\right)_0=\left(\frac{x_0^2}{2}\right)_i,\quad i=1,\dots,n-1.
\end{equation}
The conservative form of these equations allows us to write them in the form of a system as:
\begin{align}
m_0&=\frac{x_0^2}{2},\label{mcbs1}\\
m_i&=\frac{x_{i+1}}{x_0}+\frac{x_{i00}}{x_0},\quad i=1,\dots,n-1.\label{mcbs2}
\end{align}
which can be considered as  modified versions of the CBS equation with $m=m\left(T_0,..T_i,T_{i+1},...T_n\right)$. The modified CBS equation has been extensively studied from the point of view
of the Painlev\'e analysis in \cite{Sardon:estevez04}, its LP was derived and hence, a version of a LP for Qiao is available in \cite{Sardon:estevez51}.

\section{Miura-Reciprocal transformations}

As we have seen in the former section, both hierarchies CH$1+1$ and Qiao$1+1$ are related to
the CBS and mCBS respectively, through reciprocal transformations. These final equations possess the PP, whereas the initial did not accomplish it. In this way, we are able
to obtain their LPs, solutions and many other properties. The inverse reciprocal transform allows us to obtain LPs for the initial.
From the literature \cite{Sardon:estevez04}, we know that CBS and mCBS can be transformed one into another. In this manner,
the fields present in CBS and mCBS are related through the following formula:
\begin{equation}\label{miura}
4M=x_0-m.
\end{equation}
This is the point at which the question of whether Qiao$1+1$ could possibly be a modified version of CH$1+1$ arises. Nevertheless, the relation
between these two hierarchies cannot be a simple Miura transform, since each of them is written in different triples of
variables $(X,Y,T)$ and $(x,y,t)$. However, both triples lead to the same final triple $(T_0,T_1,T_n)$. Then, by combining \eqref{rtch}
and \eqref{rtq} we have:
\begin{equation}\label{mix}
PdX-\frac{1}{2}P\Omega^{(1)}dT=udx-u\omega^{(1)}dt,\quad dt=dT
\end{equation}
that yields a relationship between the variables in CH$1+1$ and Qiao$1+1$.
Using \eqref{miura}, we obtain relations between fields $X$ and $x$:
\begin{align}
4M_0&=x_{00}-m_0\Rightarrow \frac{X_{00}}{X_0}+X_0=x_0,\label{mix1}\\
4M_i&=x_{0i}-m_i\Rightarrow -\frac{X_{i+1}}{X_0}=x_{0i}-\frac{x_{00i}}{x_0}-\frac{x_{i+1}}{x_0},\quad i=1,\dots,n-1.\label{mix2}
\end{align}
where we have also employed \eqref{cbs2}, \eqref{mcbs1} and \eqref{mcbs2}.
Now, by using the inverse reciprocal transformations proposed in \eqref{irtch} and \eqref{irtq} in equations \eqref{mix1} and \eqref{mix2}, we obtain (see appendix):
\begin{equation}\label{heights}
\frac{1}{u}=\left(\frac{1}{P}\right)_X+\frac{1}{P}
\end{equation}
\begin{equation}\label{fields}
P\Omega^{(i+1)}=2(v^{(i)}-v^{(i)}_x)\Rightarrow \omega^{(i+1)}=\frac{\Omega^{(i+1)}_X+\Omega^{(i+1)}}{2},\quad i=1,\dots,n-1.
\end{equation}
Furthermore, if we isolate $dx$ in \eqref{mix} and use expression \eqref{heights}, we have:
\begin{equation}\label{dx}
dx=\left(1-\frac{P_X}{P}\right)dX+\left(\omega^{(1)}-\frac{\Omega^{(1)}}{2}\left(1-\frac{P_X}{P}\right)\right)dT.
\end{equation}
The condition $d^{2}x=0$ implies that the cross derivative satisfies:
\begin{equation}\label{crossd}
\left(1-\frac{P_X}{P}\right)_T=\left(\omega^{(1)}-\frac{\Omega^{(1)}}{2}\left(1-\frac{P_X}{P}\right)\right)_X\Rightarrow \omega^{(1)}=\frac{\Omega^{(1)}_X+\Omega^{(1)}}{2}.
\end{equation}
And finally, with the aid of \eqref{crossd}, we write \eqref{dx} like:
\begin{equation*}
dx=\left(1-\frac{P_X}{P}\right)dX-\frac{P_T}{P}dT
\end{equation*}
and can be integrated as:
\begin{equation}
x=X-\ln{P}.
\end{equation}
This equation gives us an important relation between the initial independent variables $X$ and $x$ in CH$1+1$ and Qiao$1+1$, respectively.
Notice that the rest of independent variables, $y,t$ and $Y,T$ do not appear in these expressions, since they were
left untouched in the transformations proposed in \eqref{rtch} and \eqref{rtq}.

Summarizing: hierarchies (\ref{chh}) and (\ref{qh}) are related through the Miura-reciprocal transformation 
$$x=X-\frac{\ln{U}}{2}$$
$$\frac{1}{u}=\left(\frac{1}{\sqrt U}\right)_X+\frac{1}{\sqrt U}$$
which means that the Qiao hierarchy can be considered as the modified version of the celebrated Camassa- Holm hierarchy

\section{Conclusions}

Our principal aim in this letter was to point out the utility of reciprocal transformations for the identification
of integrable equations, with the compromise of studying their properties, as well as to classify disguised versions of a unique equation. To clarify
this, we have proceeded presenting a particular example: the Camassa-Holm and Qiao hierarchies in $1+1$ dimensions for
which we have derived equations relating the fields and independent variables present in the first to the ones in the second.
For this issue, we have made use of a combination of a Miura and reciprocal transforms, hence the name {\bf Miura-reciprocal transformations}.

\section*{Acknowledgements}
This research has been supported in part by the DGICYT under project FIS2009-07880 and JCyL under contract GR224.

\section*{Appendix}
\begin{itemize}
\item {Method for obtaining equation \eqref{heights}}

Equation \eqref{mix1} provides:
$$x_0=X_0+\partial_0(\ln X_0);$$
if we use the fact that $X_0=\frac{1}{P}$ and $x_0=\frac{1}{u}$ we get:
$$\frac{1}{u}=\frac{1}{P}-\partial_0(\ln P),$$
and by using \eqref{rtch} we have
$$\frac{1}{u}=\frac{1}{P}-\frac{1}{P}(\ln P)_X$$
which yields \eqref{heights}.
\item {Method for obtaining  equation \eqref{fields}}

By taking $i=1..n-1$ in \eqref{mix2}, we have:
$$-\frac{X_{i+1}}{X_0}=x_{0,i}-\frac{x_{0,0,i}}{x_0}-\frac{x_{i+1}}{x_0}, \quad i=1..n-1.$$
If we use \eqref{irtch} and \eqref{irtq} the result is:
$$-\frac{P\Omega^{[i+1]}}{2}=\partial_0(\omega^{[i]})-u\partial_{00}(\omega^{[i]})-u\omega^{[i+1]}, \quad i=1..n-1$$
And now \eqref{rtq} gives us:
$$-\frac{P\Omega^{[i+1]}}{2}=\frac{\omega^{[i]}_x}{u}-\left(\frac{\omega^{[i]}_x}{u}\right)_x-u\omega^{[i+1]}, \quad i=1..n-1$$
If we use the following expressions arising from \eqref{qt1} and \eqref{qt2}
$$\omega^{[i]}_x=uv^{[i]}_x,\quad u\omega^{[i+1]}=v^{[i]}-v^{[i]}_{xx},\quad i=1..n-1,$$
we arrive at:
$$-\frac{P\Omega^{[i+1]}}{2}=v^{[i]}_x-v_i,\quad i=1..n-1$$
as  is required in \eqref{fields},
and $u\omega^{[i+1]}=v^{[i]}-v^{[i]}_{xx}, i=1..n-1$
can be written as
$$u\omega^{[i+1]}=\left(v^{[i]}-v^{[i]}_x\right)+\left(v^{[i]}_x-v^{[i]}_{xx}\right)=\left(\frac{P\Omega^{[i+1]}}{2}\right)+\left(\frac{P\Omega^{[i+1]}}{2}\right)_x,\quad i=1..n-1$$
and from \eqref{mix}, we have $\partial_x=\frac{u}{P}\partial_X$. Therefore,
$$u\omega^{[i+1]}=\left(\frac{P\Omega^{[i+1]}}{2}\right)+\frac{u}{P}\left(\frac{P\Omega^{[i+1]}}{2}\right)_X, \quad i=1..n-1$$
$$\omega^{[i+1]}=\left(\frac{P\Omega^{[i+1]}}{2u}\right)+\frac{1}{2P}\left(P_X\Omega^{[i+1]}+P\Omega^{[i+1]}_X\right), \quad i=1..n-1$$
We can eliminate $u$ with the aid of \eqref{heights}. The result is:
$$\omega^{[i+1]}=\frac{\Omega^{[i+1]}_X+\Omega^{[i+1]}}{2}, \quad i=1,\dots,n-1.$$
\end{itemize}

\end{document}